\documentclass[a4paper,12pt]{article}
\usepackage{amsmath}
\usepackage{amssymb}
\usepackage{amsfonts}
\usepackage{a4wide}
\usepackage{indentfirst}
\usepackage{cite}
\usepackage{color}
\usepackage{times}
\usepackage{txfonts}

\newcommand {\oks}[2]{{\raise0.7ex\hbox{${\scriptstyle #1}$}\!\mathord{\left/
{\vphantom{{1}{2}}}\right.\kern-\nulldelimiterspace}\!\lower0.7ex
\hbox{${\scriptstyle #2}$}}}

\begin{document}

\title{\bf High energy neutrino spin light}

\author{\bf A. E. Lobanov\thanks{E-mail: lobanov@phys.msu.ru}}
\date{}
\maketitle
\begin{center}
{\em Moscow State University, Department of Theoretical physics,
$119992$  Moscow, Russia}
\end{center}

\begin{abstract}
{The quantum theory of spin light (electromagnetic radiation
emitted  by a Dirac massive neutrino propagating in dense matter
due to the weak interaction of a neutrino with background
fermions) is developed. In contrast to the Cherenkov radiation,
this effect does not disappear even if  the medium refractive
index is assumed to be equal to unity. The formulas for the
transition rate and the total radiation power are obtained. It is
found out that radiation of photons is possible only when the sign
of the particle helicity is opposite to that of the effective
potential describing the interaction of a neutrino (antineutrino)
with the background medium. Due to the radiative self-polarization
the radiating particle can change its helicity. As a result, the
active left-handed polarized neutrino (right-handed polarized
antineutrino) converting to the state with inverse helicity can
become practically ``sterile''. Since the sign of the effective
potential depends on the neutrino flavor and the matter structure,
the spin light can change a ratio of active neutrinos of different
flavors. In the ultra relativistic approach, the radiated photons
averaged energy  is equal to one third of the initial neutrino
energy, and two thirds of the energy are carried out by the final
``sterile'' neutrinos.}
\end{abstract}

A Dirac massive neutrino has non-trivial electromagnetic
properties. In particular, it possesses non-zero magnetic moment
\cite{FujShr80}. Therefore a Dirac massive neutrino propagating in
dense matter can emit electromagnetic radiation due to the weak
interaction of a neutrino with background fermions \cite{L49,L55}.
As a result of the radiation, neutrino can change its helicity due
to the radiative self-polarization. In contrast to the Cherenkov
radiation, this effect does not disappear even if the refractive
index of the medium is assumed to be equal to unity. This
conclusion is valid for any model of neutrino interactions
breaking spatial parity. The phenomenon was called the neutrino
spin light in analogy with the effect, related with the
synchrotron radiation power depending  on the electron spin
orientation (see \cite{BTB95}).

The properties of spin light were investigated basing upon the
quasi-classical theory of radiation and self-polarization of
neutral particles \cite{L38,L51} with the use of the
Bargmann--Michel--Telegdi (BMT) equation \cite{BMT} and its
generalizations \cite{L43,L32}. This theory is valid when the
radiated photon energy is small as compared with the neutrino
energy, and this narrows the range of  astrophysical applications
of the obtained formulas.

In the present paper, the properties of spin light are
investigated  basing upon the consistent quantum theory, and this
allows the neutrino recoil in the act of radiation to be
considered for. The above mentioned restriction is eliminated in
this way.

On the other hand, the detailed analysis of the results of our
investigations shows that the features of the effect depend on the
neutrino flavor, helicity and the matter structure
\cite{L0411342}. This fact leads to the conclusion that the spin
light can initiate transformation of a neutrino from the active
state to a practically ``sterile'' state, and the inverse  process
is also possible.

When the interaction of a neutrino with the background fermions is
considered to be coherent, the propagation of a massive neutrino
in the matter is described by the Dirac equation with the
effective potential \cite{Wolf,MS}. In what follows, we restrict
our consideration to the case of a homogeneous and isotropic
medium. Then in the frameworks of the minimally extended standard
model, the form of this equation is uniquely determined by the
assumptions similar to those adopted in \cite{F}:
\begin{equation}\label{1}
  \left(i
  \hat{\partial}-\frac{1}{2}\hat{f}(1+\gamma^{5})-m_{\nu}\right)
  \varPsi_{\nu}=0.
\end{equation}
\noindent The function $f^{\mu}$ is a linear combination of
fermion currents and polarizations. The quantities with hats
denote scalar products of Dirac matrices with 4-vectors, i.e. ,
$\hat{a} \equiv \gamma^{\mu}a_{\mu}.$

If the medium is at rest and unpolarized then ${\bf f}
=0.$ The component $f^{0}$ calculated in the first order of the
perturbation theory is as follows \cite{PP,N,NR}:
\begin{equation}\label{2}
f^{0}=\sqrt{2}G_{\mathrm F}\bigg\{\sum\limits_{f}^{}
\left(I_{e\nu} + T_{3}^{(f)}-2Q^{(f)}\sin^{2}\theta_{\mathrm
W}\right)(n_{f}-n_{\bar{f}})\bigg\}.
\end{equation}
\noindent Here, $n_{f},n_{\bar{f}}$ are the number densities of
background fermions and anti-fermions, $Q^{(f)}$ is the electric
charge of the fermion and $T_{3}^{(f)}$ is the third component of
the weak isospin for the left-chiral projection of it. The
parameter \mbox{$I_{e\nu}$} is equal to unity for the interaction
of electron neutrino with electrons. In other cases $I_{e\nu}= 0.$
Summation is performed over all fermions $f$ of the background.

Let us obtain a solution of equation (\ref {1}). Since function
$f^{\mu} = {\mathrm{const}},$  equation (\ref {1}) commutes with
operators of canonical momentum $i{{\partial}_{\mu}}.$ However,
the commonly adopted choice of eigenvalues of this operator as quantum numbers
in this problem is not satisfactory.
Kinetic momentum components of a particle, related to
its group 4-velocity $u^{\mu}$ by the
relation $q^{\mu} = m_{\nu}u^{\mu},\; q^{2}= m^{2}_{\nu},$
are  more suitable to play the role of its quantum numbers. This
choice can be justified, since
it is the particle kinetic momentum that can be really observed.

The explicit form of the kinetic momentum operator for the
particle with spin is not known beforehand, and hence, in order to
find the appropriate solutions, we have to use the correspondence
principle.

It was shown in \cite{L43} that when the effects of the
neutrino weak interaction are taken into account, the Lorentz
invariant generalization of the BMT equation for spin vector $S^{\mu}$
is as follows:
\begin{equation}\label{04}
{\dot{S}^{\mu}} =2\mu_{0}
 \Big\{ \left(F^{\mu\nu}+G^{\mu\nu}\right)S_{\nu} -u^{\mu}
u_{\nu} \left(F^{\nu\lambda}+G^{\nu\lambda}\right)S_{\lambda}
\Big\} ,
\end{equation}
\noindent where
\begin{equation}\label{5}
G^{\mu \nu}= \frac{1}{2\mu_{0}}e^{\mu \nu \rho \lambda}
f_{\rho}u_{\lambda},
\end{equation}
and a dot denotes the differentiation with respect to the proper
time $\tau$.

Let us introduce the quasi-classical spin wave functions. Such
wave functions can be constructed as follows \cite{L51,L32}.
Suppose the Lorentz equation is solved, i. e. the dependence of particle
coordinates on proper time is found. Then the BMT equation
transforms to ordinary differential equation, whose resolvent
determines a one-parametric subgroup of the Lorentz group. The
quasi-classical spin wave function is represented by a spin-tensor,
whose evolution is determined by the same one-parametric
subgroup.

In the case when the effect of an external electromagnetic field
can be neglected as compared with the effect of the neutrino
interaction with the background matter, the equation for the
neutrino quasi-classical wave function $\varPsi(\tau)$ is

\begin{equation}\label{7}
  \dot{\varPsi}(\tau) =
  \,i\mu_{0}\gamma^{5}{}^{\star}G^{\mu\nu}u_{\nu}\gamma_{\mu}
  \hat{u}\,\varPsi(\tau),
\end{equation}
\noindent where $^{\star}G^{\mu\nu}=
-\oks{1}{2}e^{\mu\nu\rho\lambda}G_{\rho\lambda}$ is a  tensor dual
to $G^{\mu\nu}.$ Obviously, the quasi-classical density matrix of
a polarized neutrino takes the form
\begin{equation}\label{x11}
 \rho(\tau,\tau') =\frac{1}{2}\,U(\tau,\tau_{0}) \big(\hat
 {q}(\tau_{0})+m\big)
 \big(1-\gamma^5\hat
{S}(\tau_{0})\big)U^{-1}(\tau',\tau_{0}),
\end{equation}

\noindent where $U(\tau,\tau_{0})$ is the resolvent of equation
(\ref{7}), and the equation for ${U}(\tau,\tau_{0})$ is
\begin{equation}\label{9}
  \dot{U}(\tau,\tau_{0}) =
  \frac{i}{4}\,\gamma^{5}\left(\hat{f}\hat{u}
  -\hat{u}\hat{f}\right)
  U(\tau,\tau_{0}).
\end{equation}
We note that the operator $U(\tau,\tau_{0})$ is defined up to a
phase factor $e^{-iF(x)},$ with the derivative of the exponent
with respect to the proper time is equal to zero:
\begin{equation}\label{12}
\dot F(x)=0.
\end{equation}

Let us choose the solution of equation (\ref {1}) in the form
 \cite{L32}
\begin{equation}\label{10}
\varPsi(x)=U\left(\tau(x)\right)\varPsi_{0}(x)\,,
\end{equation}
where $\varPsi_0$ is a solution of the Dirac equation for a free
particle
  \begin{equation}\label{11}
\varPsi_0(x)=e^{-i(qx)}(\hat{q}+m_{\nu})(1-\gamma^{5}\hat{S}_{0}
)\psi^0.
\end{equation}
Here  $\psi^0$ is constant bispinor and $\varPsi_0(x)$ normalized
by the condition
$$ \bar{\varPsi}_0(x)\varPsi_0(x)= 2m_{\nu}. $$

Substitution of the expression (\ref{10}) in eq. (\ref{1}) results
in the relation
\begin{equation}\label{13}
  \left\{\hat{q}+(\hat{\partial}F)-\frac{1}{2}\hat{f}+\frac{1}{2}
  \gamma^{5}\hat{f}+
  \frac{1}{4}\,\gamma^{5}\hat{N}\left(\hat{f}\hat{u}-\hat{u}\hat{f}\right)
  -m_{\nu}\right\}e^{- iF(x)}U\left(\tau(x)\right)\varPsi_{0}=0,
\end{equation}
\noindent where $N^{\mu}=\partial^{\mu}\tau.$ Since the commutator
$[\hat{q},U]=0,$ and the matrix $U$ is nondegenerate, then for
this relation to hold the following  condition is required
\begin{equation}\label{14}
  (\hat{\partial}F)-\frac{1}{2}\hat{f}+\frac{1}{2}
  \gamma^{5}\hat{f}+
  \frac{1}{4}\,\gamma^{5}\hat{N}\left(\hat{f}\hat{u}-\hat{u}\hat{f}\right)
  =0.
\end{equation}
\noindent It is easy to find out that the equation (\ref{14}) is
valid only if
\begin{equation}\label{15}
\begin{array}{l}
 \displaystyle
 {\partial}^{\lambda}F=\frac{1}{2}{f}^{\lambda},\quad
 \displaystyle  e^{\mu\nu\rho\lambda}N_{\mu}f_{\nu}u_{\rho}=0, \quad
 \displaystyle  \left(1-(Nu)\right)f^{\lambda}+(Nf)u^{\lambda}=0.
\end{array}
\end{equation}
\noindent From two latter equations it  follows that
\begin{equation}\label{16}
  N^{\mu}=\frac{f^{\mu}(fu)-u^{\mu}f^{2}}{(fu)^{2}-f^{2}u^{2}}.
\end{equation}

So $f^{\mu} = {\mathrm {const}},$ then
\begin{equation}\label{17}
  \tau = (Nx),\qquad F = \frac{1}{2} (fx),
\end{equation}
\noindent and we can write
\begin{equation}\label{18}
  U(x) = e^{{-i}(fx)/{2}}\sum_{\zeta = \pm 1}
  e^{i\zeta\varphi}\Lambda_{\zeta}.
\end{equation}
\noindent Here
\begin{equation}\label{19}
  \Lambda_{\zeta} = \frac{1}{2}
  \left[1-
  \zeta\gamma^{5}\hat{S}_{tp}\hat{q}/m_{\nu}\right],\qquad
  \zeta \pm 1,
\end{equation}
\noindent are spin projection operators with eigenvalues $\zeta
\pm 1$ respectively, and
\begin{equation}\label{20}
\varphi =
\frac{\tau}{2}\sqrt{(fq)^{2}-f^{2}m^{2}_{\nu}}=\frac{(fq)(fx)
  - f^{2}(qx)}{2\sqrt{(fq)^{2}-f^{2}m^{2}_{\nu}}},\quad\quad
  {S}_{tp}^{\mu}= \frac{q^{\mu}(fq)/m_{\nu}
  - f^{\mu}m_{\nu}}{\sqrt{(fq)^{2}-f^{2}m^{2}_{\nu}}}.
\end{equation}

From the obtained formulas it follows that the eigenvalues of the
operator of canonical momentum $i\partial^{\mu}$ are
\begin{equation}\label{a20} P^{\mu}= q^{\mu}\left(1+\frac{\zeta
f^{2}}{2\sqrt{(fq)^{2}-f^{2}m^{2}_{\nu}}}\right)
+\frac{f^{\mu}}{2}\left(1-\frac{\zeta
(fq)}{\sqrt{(fq)^{2}-f^{2}m^{2}_{\nu}}}\right).
\end{equation}
The dispersion law follows from eq. (\ref{a20}) in the form
\begin{equation}\label{b20}
P^{2}= m_{\nu}^{2}+(Pf)-{f^{2}}/{2}- \zeta
\sqrt{\left((Pf)-{f^{2}}/{2}\right)^{2}-f^{2}m^{2}_{\nu}}.
\end{equation}

If the medium is at rest  and unpolarized then the neutrino total
energy and canonical momentum are determined by the formulas
\begin{equation}\label{021}
\varepsilon = q^{0} + f^{0}/2, \qquad {\bf{P}} = {\bf q}\Delta_{q
\zeta},
\end{equation}
\noindent where $\Delta_{q \zeta} = 1+\zeta f^{0}/2|{\bf{q}}|,$
and
\begin{equation}\label{0021}
S_{tp}^{\mu}=\frac{1}{m_{\nu}}\left\{|{\bf q}|,
 q^{0}{\bf q}/|{\bf q}|\right\},
\end{equation}
\noindent i. e. the eigenvalues $\zeta = \pm 1$ determine the
helicity of the particle. Consequently,  the dispersion law is
\begin{equation}\label{22}
\varepsilon = \sqrt{\left(\Delta |{\bf{P}}|- \zeta
f^{0}/2\right)^{2}+m^{2}_{\nu}} + f^{0}/2,
\end{equation}
\noindent where $\Delta = {\mathrm{sign}\left(\Delta_{q \zeta}
\right)}.$ Obviously
$$\frac{\partial \varepsilon}{\partial {\bf{P}}}=\frac{\bf{q}}{q^{0}}$$
\noindent is  the particle group velocity.

The relation (\ref{22}) differs those used in previous papers
(see, for example,\cite{P}) by the multiplier $\Delta.$ This is
due to the fact that, in these papers the projection of the
particle spin on the canonical momentum ${\bf{P}}$ and not the
helicity of the particle was used as the spin quantum number
$\zeta$.  The helicity is the projection of the spin on the
direction of its kinetic momentum \cite{ST,BG,T}, because the rest
frame of the particle is determined by the condition that its
group velocity is equal to zero. In our problem the directions of
canonical and kinetic momenta, generally speaking, are different,
and hence,  the projection of particle spin on the canonical
momentum does not coincide with its helicity.

From formulas (\ref{021}), it is seen that if the energy is
expressed in terms of the kinetic momentum, then it does not
depend on the particle helicity, while the particle canonical
momentum is a function of the helicity. Therefore, the statement
of the authors of \cite{ST0410297}, i.e., that the radiation of
photons in the process of the spin light emission takes place due
to neutrino transitions from the ``exited'' helicity state to the
low-lying helicity  state in matter,  is not correct.

Let us consider the process of emitting photons by a massive
neutrino in  unpolarized matter at rest. In this case, the
orthonormalized system of solutions for equation (\ref{1}) is:
\begin{equation}\label{21}
\varPsi(x)= \frac{\left|\Delta_{q \zeta}\right|}{\sqrt{2q^{0}}}
\,e^{-i(q^{0}+f^{0}/2)x^{0}} e^{i{\bf{qx}}\Delta_{q \zeta} }
(\hat{q}+m_{\nu})(1-\zeta\gamma^{5}\hat{S}_{tp} )\psi^0.
\end{equation}

The formula for the spontaneous radiation transition probability
of a neutral fer\-mion with anomalous magnetic moment $\mu_{0}$
is{\footnote{In the expression for the radiation energy $\mathcal
E,$ the additional factor $k$, i.e. the energy of radiated photon,
appears in the integrand.}}:
\begin{equation}\label{ar1}
\begin{array}{c}
  \displaystyle P=-\frac{1}{2p^{0}}\!\int\!\! d^{4}x\,d^{4}y\!\int
  \frac{d^{4}q\,d^{4}k}{(2\pi)^{6}}\,
  \delta(k^{2})\delta(q^{2}\!-m^{2}_{\nu})\times
  \\[8pt]
\displaystyle \times {\mathrm{Sp}}
\big\{\varGamma_{\mu}(x)\varrho_{i}(x,y;p,\zeta_{i})
\varGamma_{\nu}(y)\varrho_{f}(y,x;q,\zeta_{f})\big\}
\varrho^{\mu\nu}_{ph}(x,y;k).
\end{array}
\end{equation}
\noindent \noindent Here,
$\varrho_{i}(x,y;p),\;\varrho_{f}(y,x;q)$ are density matrices of the
initial $(i)$ and final $(f)$ states of the fermion,
$\,\varrho^{\mu\nu}_{ph}(x,y;k)$ is the radiated photon density
matrix, $\varGamma^{\mu} = -\,\sqrt{4\pi}
\mu_{0}\sigma^{\mu\nu}k_{\nu}$ is the vertex function. The density
matrix of longitudinally polarized neutrino in the unpolarized
matter at rest constructed with the use of the solutions of
equation (\ref{1}) has the form
\begin{equation}\label{arx4}
\varrho(x,y;p,\zeta)=\frac{1}{2}\Delta_{p\zeta}^{2}(\hat{p}+m_{\nu})
(1-\zeta\gamma^{5}\hat{S}_{
p})e^{-i(x^{0}-y^{0})(p^{0}+f^{0}/2)+i({\bf x}-{\bf y}) {\bf
p}\Delta_{p\zeta}}.
\end{equation}

After summing over photon polarizations{\footnote{We do not
consider the polarization of spin light photons here. In the
quasi-classical approximation, this problem was investigated in
\cite{ST0410297}.}} and integrating with respect to coordi\-nates
we obtain the expression for the transition rate under
investi\-ga\-tion:
\begin{equation}\label{arxx4}
\begin{array}{c}
\displaystyle W= \frac{\mu^{2}_{0}}{p^{0}}\!\!\int
  \frac{d^{4}q\,d^{4}k}{(2\pi)}\,
  \delta(k^{2})\delta(q^{2}\!-m^{2}_{\nu})\delta(p^{0}-q^{0}-k^{0})
  \delta^{3}
  ({\bf p}\Delta_{p\zeta_{i}}-{\bf q}\Delta_{q\zeta_{f}}-{\bf
  k})T(p,q),
\end{array}
\end{equation}
\noindent where
\begin{equation}\label{arxxx4}
\begin{array}{c}
T(p,q)=4\Delta_{p\zeta_{i}}^{2}\Delta_{q\zeta_{f}}^{2}
\left\{(pk)(qk)-\zeta_{i}\zeta_{f}\left[k^{0}|{\bf p}|-p^{0}({\bf
  p}{\bf
  k})/|{\bf p}|\right]\left[k^{0}|{\bf q}|-q^{0}({\bf
  q}{\bf
  k})/|{\bf q}|\right]\right\}.
\end{array}
\end{equation}

After integrating over  ${\bf k},$ $k^{0},$ $|{\bf q}|$
we obtain the spectral-angular distribution of the final neutrino
\begin{equation}\label{q3}
\begin{array}{c}
\displaystyle W=
-\zeta_{i}\zeta_{f}\frac{\mu^{2}_{0}}{\pi{p^{0}{|\bf
p|}}}\!\!\int\limits_{m_{\nu}}^{p^{0}}{dq^{0}}\Delta_{p\zeta_{i}}
\Delta_{q\zeta_{f}}\int\!\! dO
\,\times\\[12pt]
\displaystyle\times\delta\left((p^{0}-q^{0})^{2}+2|{\bf p}||{\bf
q}|\Delta_{p\zeta_{i}}\Delta_{q\zeta_{f}}\cos\vartheta_{\nu}-|{\bf
p}|^{2}\Delta_{p\zeta_{i}}^{2}-|{\bf
q}|^{2}\Delta_{q\zeta_{f}}^{2}\right)\times\\[12pt]
\displaystyle \!\!\!\!\times
\!\left\{\!(f^{0}/2)^{2}\!\left[\zeta_{f}|{\bf p}||{\bf
q}|+\zeta_{i}(m^{2}_{\nu}-p^{0}q^{0})\right]^{2}\!\!+\!
\left[(f^{0}/2)(\zeta_{i}q^{0}|{\bf p}|-\zeta_{f}p^{0}|{\bf
q}|)+m^{2}_{\nu}(p^{0}-q^{0})\right]^{2}\!\right\}\!,\end{array}
\end{equation}
\noindent where $$|{\bf q}|=\sqrt{(q^{0})^{2}-m^{2}_{\nu}}.$$

It is convenient to express the results  of integrating over
angular variables using dimensionless quantities.
Introducing the notations
\begin{equation}\label{q10}
  x=q^{0}/m_{\nu},\quad \gamma=p^{0}/m_{\nu},\quad
  d=|f^{0}|/2m_{\nu},\quad
  \bar{\zeta}_{i,f}={\zeta}_{i,f}\,{\mathrm{sign}}(f^{0})
\end{equation}
we have
\begin{equation}\label{q16}
\begin{array}{c}
\displaystyle W_{\bar{\zeta}_{f}}=
\frac{\mu_{0}^{2}m_{\nu}^{3}}{\gamma(\gamma^{2}-1)}\int
  \frac{dx}{\sqrt{x^{2}-1}}\bigg\{d^{2}\left[\bar{\zeta}_{f}
  {\sqrt{\gamma^{2}-1}}{\sqrt{x^{2}-1}}-\bar{\zeta}_{i}(\gamma x
  -1)\right]^{2}+\\[12pt]+\left[\gamma-x +
  d\left(\bar{\zeta}_{i}x{\sqrt{\gamma^{2}-1}}-\bar{\zeta}_{f}
  \gamma\sqrt{x^{2}-1}\right)\right]^{2}
  \bigg\}.
\end{array}
\end{equation}
\noindent The integration bounds  in the formula (\ref{q16})
are
\begin{equation}\label{q011}
\begin{array}{ll}
x\in \emptyset & \quad \gamma \in [1,\infty),
\end{array}
\end{equation}
\noindent if $\bar{\zeta}_{i}=1,$
\begin{equation}\label{q11}
\begin{array}{ll}
x\in \emptyset & \quad \gamma \in [1,\gamma_{0}),\\
x\in[\omega_{1},\omega_{2}] & \quad \gamma \in
[\gamma_{0},\gamma_{1}),\\  x\in[1,\omega_{2}] & \quad \gamma \in
[\gamma_{1},\gamma_{2}),\\  x\in \emptyset &
\quad \gamma \in [\gamma_{2},\infty),\\
\end{array}
\end{equation}
\noindent if $\bar{\zeta}_{i}=-1,\bar{\zeta}_{f}= -1,$ and
\begin{equation}\label{q12}
\begin{array}{ll}
x\in \emptyset &
\quad \gamma \in [1,\gamma_{1}),\\
x\in[1,\omega_{1}] &
\quad \gamma \in [\gamma_{1},\gamma_{2}),\\
x\in[\omega_{2},\omega_{1}] &
\quad \gamma \in [\gamma_{2},\infty).\\
\end{array}
\end{equation}
\noindent if $\bar{\zeta}_{i}=-1,\bar{\zeta}_{f}= 1.$

\noindent Here
\begin{equation}\label{q13}
\omega_{1}= \frac{1}{2}\left(z_{1} + z_{1}^{-1}\right),\qquad
\omega_{2}= \frac{1}{2}\left(z_{2} + z_{2}^{-1}\right),
\end{equation}
where
\begin{equation}\label{q14}
\begin{array}{l}
z_{1}=\displaystyle \gamma+\sqrt{\gamma^{2}-1}-2d,\\[8pt]
z_{2}=\displaystyle \gamma-\sqrt{\gamma^{2}-1}+2d,
\end{array}
\end{equation}
and
\begin{equation}\label{q15}
\begin{array}{ll}
\gamma_{0}=\displaystyle \sqrt{1+d^{2}}, & {}\\[8pt]
\displaystyle \gamma_{1}=\frac{1}{2}\left\{\left(1+2d\right)
 +\left(1+2d\right)^{-1}\right\},& {}\\[12pt]
 \displaystyle \gamma_{2}=\frac{1}{2}\left\{\left(1-2d\right)
 +\left(1-2d\right)^{-1}\right\}& d< 1/2,\\[16pt]
 \displaystyle \gamma_{2}= \infty & d\geqslant 1/2.
\end{array}
\end{equation}

The integration is carried out elementary. The transition rate
under investi\-ga\-tion is defined as
\begin{equation}\label{q18}
\begin{array}{c}
\displaystyle W_{\bar{\zeta}_{f}}=\frac{\mu_{0}^{2}m_{\nu}^{3}}{4}
\Big\{(1+ \bar{\zeta}_{f})\left[Z(z_{1},1)
  \Theta(\gamma -\gamma_{1})
  +Z(z_{2},-1)\Theta(\gamma -\gamma_{2})\right]+\\[12pt]
\displaystyle +(1-\bar{\zeta}_{f})\left[Z(z_{1},1)
  \Theta(\gamma_{1} -\gamma)
  +Z(z_{2},-1)\Theta(\gamma_{2} -\gamma)\right]
  \Theta(\gamma -\gamma_{0})\Big\}(1-\bar{\zeta}_{i}).
\end{array}
\end{equation}
\noindent Here
\begin{equation}\label{q17}
\begin{array}{c}
\displaystyle
Z(z,\bar{\zeta}_{f})=\frac{1}{\gamma(\gamma^{2}-1)}\left\{\ln
z\left[\gamma^{2}
+d\sqrt{\gamma^{2}-1}+d^{2}+{1}/{2}\right]\right.+\\[12pt]
\displaystyle +\frac{1}{4}\left(z^{2}-z^{-2}\right)
\left[d^{2}\left(2\gamma^{2}-1\right)
+d\sqrt{\gamma^{2}-1}+{1}/{2}\right]+\\[12pt]
\displaystyle +\frac{\bar{\zeta}_{f}}{4}\left(z -z^{-1}\right)^{2}
\left[2d\sqrt{\gamma^{2}-1}+1\right]d\gamma-\\[12pt]
\displaystyle -\left(z  -z^{-1}\right)
\left[d^{2}  + d\sqrt{\gamma^{2}-1}+1\right]\gamma-\\[12pt]
\displaystyle -\left.\bar{\zeta}_{f}\left(z+z^{-1} -2\right)
\left[d\sqrt{\gamma^{2}-1}+\gamma^{2}\right]d\right\}.
\end{array}
\end{equation}
Therefore, the transition rate after summation  over polarizations
of the final neutrino becomes
\begin{equation}\label{q19}
W_{\bar{\zeta}_{f}=1}+W_{\bar{\zeta}_{f}=-1}=
\frac{{\mu_{0}^{2}m_{\nu}^{3}}}{2}(1-\bar{\zeta}_{i})\Big\{Z(z_{1},1)
+Z(z_{2},-1)\Big\} \Theta(\gamma -\gamma_{0}).
\end{equation}

If $d\gamma \ll 1,$ then expression (\ref{q18}) leads to the
formula
\begin{equation}\label{q20}
W_{\bar{\zeta}_{f}}=\frac{16\mu_{0}^{2}m_{\nu}^{3}d^{3}}{3\gamma}
(\gamma^{2}-1)^{3/2}(1-\bar{\zeta}_{i})(1+\bar{\zeta}_{f}),
\end{equation}
obtained in the quasi-classical approximation in \cite{L55}.

In the ultra-relativistic limit $(\gamma \gg 1, \; d\gamma \gg
1),$ the transition rate is given by the expression
\begin{equation}\label{q21}
W_{\bar{\zeta}_{f}}=\mu_{0}^{2}m_{\nu}^{3}d^{2}{\gamma}
(1-\bar{\zeta}_{i})(1+\bar{\zeta}_{f}).
\end{equation}

Let us consider now the radiation power. If we introduce the
function
\begin{equation}\label{q22}
  \tilde{Z}(z,\bar{\zeta}_{f}) = \gamma Z(z,\bar{\zeta}_{f})
   - Y(z,\bar{\zeta}_{f}),
\end{equation}
\noindent where
\begin{equation}\label{q23}
\begin{array}{c}
\displaystyle
Y(z,\bar{\zeta}_{f})=\frac{1}{\gamma(\gamma^{2}-1)}\left\{-\ln
z\left[d^{2}
+d\sqrt{\gamma^{2}-1}+{1}\right]\gamma\right.-\\[12pt]
\displaystyle -\frac{1}{4}\left(z^{2}-z^{-2}\right) \left[d^{2}
+d\sqrt{\gamma^{2}-1}+{1}\right]\gamma+\\[12pt]
\displaystyle +\frac{1}{12}\left(z-z^{-1}\right)^{3}
\left[d^{2}\left(2\gamma^{2}-1\right)
+d\sqrt{\gamma^{2}-1}+{1}/{2}\right]+\\[12pt]
\displaystyle +
\frac{1}{2}\left(z-z^{-1}\right)\left[2d^{2}\gamma^{2}
+2d\sqrt{\gamma^{2}-1} + \gamma^{2} +{1}\right]+\\[12pt]
\displaystyle
+\frac{\bar{\zeta}_{f}}{12}\left(\left(z+z^{-1}\right)^{3}-8\right)
\left[2d\sqrt{\gamma^{2}-1}+1\right]d\gamma -\\[12pt]
\displaystyle -\left.\frac{\bar{\zeta}_{f}}{4}\left(z -z^{-1}
\right)^{2}\left[d\sqrt{\gamma^{2}-1}+\gamma^{2}\right]d \right\},
\end{array}
\end{equation}
\noindent then the formula for the total radiation power can be
obtained from (\ref{q18}), (\ref{q19}) by the substitution
${Z}(z,\bar{\zeta}_{f}) \rightarrow \tilde{Z}(z,\bar{\zeta}_{f}).$
It can be verified that if  $d\gamma \ll 1$ then the radiation
power is
\begin{equation}\label{q24}
I_{\bar{\zeta}_{f}}=\frac{32\mu_{0}^{2}m_{\nu}^{4}d^{4}}{3}
(\gamma^{2}-1)^{2}(1-\bar{\zeta}_{i})(1+\bar{\zeta}_{f}).
\end{equation}
\noindent This result was obtained in the quasi-classical
approximation in \cite{L49}. In the ultra-relativistic limit, the
radiation power is equal to
\begin{equation}\label{q25}
I_{\bar{\zeta}_{f}}=\frac{1}{3}\mu_{0}^{2}m_{\nu}^{4}d^{2}{\gamma}^{2}
(1-\bar{\zeta}_{i})(1+\bar{\zeta}_{f}).
\end{equation}
\noindent It can be seen from equations (\ref{q21}) and
(\ref{q25}) that in the ultra-relativistic limit the averaged
energy of emitted photons is $\langle \varepsilon_{\gamma} \rangle
= \varepsilon_{\nu}/3.$ It should be pointed out that the obtained
formulas are valid both for a neutrino and for an antineutrino.
The charge conjugation operation leads to the change of the sign
of the effective potential and the replacement of the left-hand
projector by the right-hand one in the equation (\ref{1}). Thus
the sign in front of the $\gamma^{5}$ matrix remains invariant.

Using eq. (\ref{arxx4}), it is possible to find the dependence of the
radiated photon energy on the angle $\vartheta_{\gamma}$ between
the direction of the neutrino propagation and the photon wave
vector:
\begin{equation}\label{q025}
  \frac{k^{0}}{m_{\nu}}=2d\frac{\beta X -d/\gamma}{\left(X+d/\gamma\right)
  \left(X-d/\gamma\right)}.
\end{equation}
Here $\beta = \sqrt{\gamma^{2}-1}/\gamma$ is the neutrino velocity
and $X = 1-\left(\beta-d/\gamma\right)\cos \vartheta_{\gamma}.$ In
the quasi-classical approximation, this formula reduces to the
relation
\begin{equation}\label{q0025}
  \frac{k^{0}}{m_{\nu}}=\frac{2d\beta }{1-\beta
  \cos \vartheta_{\gamma}},
\end{equation}
which  follows from the results of \cite{L55} after Lorentz
transformation to the laboratory frame.

The following conclusions can be made from the obtained results. A
neutrino (anti\-neutrino) can emit photons due to coherent
interaction with  matter only  when its helicity has the sign
opposite to the sign of the effective potential $f^{0}.$
Otherwise, radiation transitions are impossible. In the case of
low energies of the initial neutrino, only  radiation without
spin-flip is possible and the probability of the process is very
low. At high energies, the main contribution to radiation is given
by the transitions with the spin-flip, the transitions without
spin-flip are either absent  or their probability is negligible.
This leads to total self-polarization, i.~e. the initially
left-handed polarized neutrino (right-handed polarized
antineutrino) is transformed to practically ``sterile''
right-handed polarized neu\-trino (left-handed polarized
antineutrino). For ``sterile'' particles, the situation is
opposite.  They can be converted to the active  form in the medium
``transparent'' for the active neutrino.

With the use of the effective potential calculated in the first
order of the perturbation theory (\ref{2}), the following
conclusions can be made. If the matter consists only of  electrons
then, in the framework of the minimally extended standard model in
the ultra-relativistic limit (here we use gaussian units), we have
for the transition rate
\begin{equation}\label{q29}
\displaystyle
W_{\bar{\zeta}_{f}}=\frac{\alpha\varepsilon_{\nu}}{32\,\hbar}
\left(\frac{\mu_{0}}{\mu_{\mathrm B}}\right)^{2}
\left(\frac{\tilde{G}_{{\mathrm F}}\,n_{e}}{m_{e}c^{2}}\right)^{2}
(1-\bar{\zeta}_{i})(1+\bar{\zeta}_{f}),
\end{equation}
and for the total radiation power
\begin{equation}\label{q30}
\displaystyle
I_{\bar{\zeta}_{f}}=\frac{\alpha\varepsilon_{\nu}^{2}}{96\,\hbar}
\left(\frac{\mu_{0}}{\mu_{\mathrm B}}\right)^{2}
\left(\frac{\tilde{G}_{{\mathrm F}}\,n_{e}}{m_{e}c^{2}}\right)^{2}
 (1-\bar{\zeta}_{i})(1+\bar{\zeta}_{f}).
\end{equation}
\noindent Here $\varepsilon_{\nu}$ is the neutrino energy,
$\mu_{\mathrm B}=e/2m_{e}$ is the Bohr magneton, $\alpha $ is the
fine structure constant, $m_{e}$ is the electron mass and
$\tilde{G}_{{\mathrm F}}={G}_{{\mathrm
F}}(1+4\sin^{2}\theta_{\mathrm W}),$ where ${G}_{{\mathrm F}},\,
\theta_{\mathrm W}$ are the Fermi constant and the Weinberg angle
respectively. Thus, after the radiative transition, two thirds of
the initial active neutrino energy are carried away by the final
``sterile'' one.

At the same time, as it can be seen from eq. (\ref{2}), a muon
neutrino in the electron medium does not emit any radiation.
Moreover, a muon neutrino does not emit radiation in an
electrically neutral medium, when the number density of protons is
equal to the electron number density. And an electron neutrino can
emit radiation if the electron number density is greater than the
neutron number density. An example of such medium is provided by
the Sun. Therefore the spin light can change the ratio of active
neutrino of different flavors.

It is obviously that the above conclusions change to opposite if
the matter consists of antiparticles. Therefore the neutrino spin
light can serve as a tool for determination of the type of
astrophysical objects, since neutrino radiative transitions in
dense matter can result in radiation of photons of super-high
energies, even exceeding the GZK cutoff. Indeed, the neutron
medium is ``transparent'' for all active neutrinos, but an active
antineutrino emits radiation in such a medium,  the transition
rate and the total radiation power can be obtained from equations
(\ref{q29}) and (\ref{q30}) after substitution
$\tilde{G}_{{\mathrm F}} \rightarrow {G}_{{\mathrm F}}, n_{e}
\rightarrow n_{n}.$ If the density of the neutron star is assume
to be $n \approx 10^{38},$ the transition rate is estimated as
\begin{equation}\label{q31}
\displaystyle W = 10^{22}\frac{
\varepsilon_{\nu}}{\varepsilon_{{\mathrm GZK}}}
\left(\frac{\mu_{0}}{\mu_{\mathrm B}}\right)^{2},
\end{equation}
where $\varepsilon_{{\mathrm GZK}} = 5\times 10^{19}$eV is GZK
cutoff energy. Although the transition rate determined by
eq. (\ref{q31}) is extremely low, this effect can still serve as one
of a possible explanations of the cosmic ray paradox.

The spin light can also be important  for the understanding of the
``dark matter'' formation mechanism in the early stages of
evolution of the Universe.

When the present paper was already submitted for publication, we
came across an article \cite{{GST0502231}}, where the spin light
theory was also considered.  The formulas of \cite{{GST0502231}}
in the ultra-relativistic limit of physical interest reproduce the
results for the  transition rate and the total power of spin light
already obtained in our earlier publication \cite{L0411342}.

\bigskip

The author is grateful to V.G.~Bagrov, A.V.~Borisov, and
V.Ch.~Zhukovsky for fruitful discussions.

\bigskip

This work was supported in part by the grant of President of
Russian Federation for leading scientific schools (Grant SS ---
2027.2003.2).

\end{document}